\documentclass[letter]{aa}          % for the letters 
\usepackage{graphicx}
\usepackage{txfonts}
\newcommand{\myrule}{\rule[-0.1cm]{0.cm}{0.5cm}} %gr\"osserer zeilenabstand
\newcommand\lsun{L_{\odot}}
\newcommand\msun{M_{\odot}}

\newcommand\mjup{M_\mathrm{Jup}}

\newcommand\mperyr{M_{\odot}\,\mathrm{yr}^{-1}}
\newcommand\kms{km\,s$^{-1}$}
\newcommand\ots{OTS\,}

\begin{document}

   \title{\ots44: Disk and accretion at the planetary border\thanks{Based 
          on observations at the Very Large Telescope of the 
	  European Southern Observatory at Paranal, Chile 
	  in program 80.C-0590(A)
	  }}

\titlerunning{\ots44: Disk and accretion at the planetary border}

   \author{V. Joergens,\inst{1,2}
           M. Bonnefoy, \inst{1}
           Y. Liu, \inst{3,4}
           A. Bayo, \inst{1,5}
           S. Wolf, \inst{4}
           G. Chauvin, \inst{6}
           P. Rojo \inst{7}
          }

   \institute{
	     Max-Planck Institut f\"ur Astronomie, 
             K\"onigstuhl~17, 69117 Heidelberg, Germany,
             \email{viki@mpia.de}
	\and  
        Universit\"at Heidelberg,
	Zentrum f\"ur Astronomie,
	Inst. f\"ur Theor. Astrophysik,
	Albert-Ueberle-Str. 2,
	69120 Heidelberg, Germany
        \and
        Purple Mountain Observatory \& Key Laboratory for Radio Astronomy, 
        Chinese Academy of Sciences, Nanjing 210008, China
        \and
        Institut f\"ur Theoretische Physik und Astrophysik,
	Universit\"at Kiel,  
        Leibnizstr. 15, 24118 Kiel, Germany
        \and
        European Southern Observatory, Alonso de C\'ordova 3107, Vitacura, Santiago, Chile
        \and
        Inst. of Planetology \& Astrophysics Grenoble,
        414, Rue de la Piscine, Domaine Universitaire, 
        38400 Saint-Martin d'H\'eres, France
        \and
        Departamento de Astronomia, Universidad de Chile, Casilla 36-D, Santiago, Chile
             }

   \date{Received; accepted}

  \abstract
   {We discover that the very low-mass brown dwarf \ots44 
(M9.5, $\sim$12\,M$_{\rm{Jup}}$) has significant accretion and a substantial disk,
which demonstrates that the processes that accompany canonical star formation
occur down to a central mass of a few Jupiter masses.
We discover in VLT/SINFONI spectra that \ots44 
has strong, broad, and variable Pa\,$\beta$ emission 
that is evidence for active accretion at the planetary border.
We also detect strong H$\alpha$ emission of \ots44 in a literature spectrum 
and determine an H$\alpha$ EW (-141\,{\AA}) that is indicative of active accretion.
Both the Pa\,$\beta$ and H$\alpha$ emission lines have broad profiles 
with wings extending to velocities of about $\pm$200\,\kms.
We determine the mass accretion rate of \ots44 based on H$\alpha$ 
to 7.6$\times 10^{-12}$\,$\mperyr$, which shows that \ots44 has a relatively high 
mass-accretion rate considering its small central mass.
This mass rate is nevertheless consistent with the general decreasing trend found 
for stars of several solar masses down to brown dwarfs.
Furthermore, we determine the properties of the disk surrounding \ots44 through 
radiative transfer modeling of flux measurement from the optical to the far-IR (Herschel)
by applying a Bayesian analysis. 
We find that \ots44 has a highly flared disk ($\beta>$1.2) with a mass of 
9.1\,$^{+1.7}_{-5.5}\times10^{-5}\,\msun$ ($\sim$0.1\,$\mjup$ or $30\,M_{Earth}$). 
We show that the ratio of disk-to-central-mass of about 10$^{-2}$ found for 
objects between 0.03\,$\msun$ and 14\,$\msun$ is also valid for \ots44 at a mass of $\sim$0.01\,$\msun$.
Our observations are in line with an isolated star-like mode of the formation
of brown dwarfs down to 0.01\,$\msun$.
   }
   \keywords{brown dwarfs -
             Stars: pre-main sequence -
             circumstellar matter -
             Accretion -
             Stars: formation -
	     stars: individual (\mbox{OTS\,44})
               }

\authorrunning{Joergens et al.}
\maketitle

%
%________________________________________________________________

\section{Introduction}
\label{sect:intro}

One of the main open questions in the theory of star formation is: 
How do brown dwarfs form. 
A high-density phase is necessary for the gravitational 
fragmentation to create very small Jeans-unstable cores. 
Proposed scenarios to prevent a substellar core in a dense environment
from accreting to stellar mass are (i) ejection
of the core through dynamical interactions, 
(ii) photo-evaporation of the gas envelope through radiation of a nearby hot star, 
and (iii) disk instabilities in circumstellar disks. 
Alternatively, brown dwarfs could form in an isolated mode by direct collapse. 
For example, filament collapse (e.g., Inutsuka \& Miyama 1992) might form
low-mass cores that experience high self-erosion
in outflows and become brown dwarfs (Machida et al. 2009).
A key to understanding star and brown dwarf formation is to observationally define 
the minimum mass that the star formation process can produce by detecting and 
exploring the main features characteristic of star formation,
such as disks, accretion, and outflows, for very low-mass objects.

Young brown dwarfs were shown to have substantial circumstellar disks 
at far-infrared (far-IR) and mm wavelengths (e.g., Harvey et al. 2012; Ricci et al. 2013).
Many of these brown dwarfs were found to actively accrete material through the disk
onto the central object
(e.g., Rigliaco et al. 2012),
and a handful of them also to drive outflows 
(e.g., Whelan et al. 2005; Phan-Bao et al. 2008;
Bacciotti et al. 2011; Joergens et al. 2012a, 2012b;
Monin et al. 2013). 
Among the lowest-mass isolated objects found to harbor a disk are
Cha\,110913-773444 ($\sim$8\,$\mjup$, Luhman et al. 2005a), 
LOri\,156 ($\sim$23\,$\mjup$, Bayo et al. 2012),
and \ots44 ($\sim$12\,M$_{\rm{Jup}}$, Luhman et al. 2005b). 

\ots44, which is the subject of the present work,
was first identified as a brown dwarf candidate in a deep near-IR imaging survey 
in the Chamaeleon~I 
star-forming region (Oasa et al. 1999). It was confirmed to be a 
very low-mass brown dwarf of spectral-type M9.5 with an estimated mass of $\sim$15\,M$_{\rm{Jup}}$
based on low-resolution near-IR and optical spectra 
(Luhman et al. 2004; Luhman 2007).
Bonnefoy et al. (2013) recently confirmed in a near-IR study a mass 
in the planetary regime ($\sim$6-17\,M$_{\rm{Jup}}$).
Mid- and far-IR excess emission detected with Spitzer (Luhman et al. 2005b) 
and Herschel (Harvey et al. 2012) indicated that \ots44 has a disk.
We present SINFONI / VLT spectroscopy of \ots44 that reveals strong 
Paschen\,$\beta$ emission, an analysis of H$\alpha$ emission in a spectrum from Luhman (2007), and 
a detailed modeling of the spectral energy distribution (SED) of the disk
of \ots44 based on Herschel data.

%__________________________________________________________________
\section{Observations}
\label{sect:obs}

{\bf The spectral energy distribution.}
To model the disk and photosphere of \ots44, we compiled optical to far-IR flux 
measurements from the literature. 
In the optical, we use I-band data from Luhman et al. (2005b) 
and an R-band magnitude of 23.5$\pm$0.1 (K. Luhman, pers. comm.).
The near-IR (JHK) regime is covered by 
observations of 2MASS 
and WISE. 
We note that the WISE W4 photometry of \ots44 was not used because of
contamination by a bright spike.
Mid-IR photometry of OTS44 was obtained by 
Spitzer using 
IRAC (3.6, 4.5, 5.8, $8.0\,\mu\rm{m}$) and 
MIPS ($24\,\mu\rm{m}$,
Luhman et al. 2008). 
Recently, OTS44 was 
observed in the far-IR (70, $160\,\mu\rm{m}$) by 
PACS/Herschel
(Harvey et al. 2012).
 
{\bf Near-infrared SINFONI integral field spectroscopy.} 
We observed \ots44 with the {\it Spectrograph for INtegral Field Observations in the Near Infrared} 
(SINFONI) 
at the VLT on December~14 and 21, 2007. 
The observations were conducted as part of a program designed to provide a library of 
near-IR spectra of young late-type objects (Bonnefoy et al. 2013). 
The instrument was operated with pre-optics and gratings enabling medium-resolution 
(R=$\lambda$/$\Delta \lambda$$\sim$2000) J-band spectroscopy (1.1-1.4 $\mu$m) with a spatial sampling of 
125$\times$250 mas/pixel. 
We reduced the data with the SINFONI data reduction pipeline version 1.9.8 
and custom routines (cf. Bonnefoy et al. 2013 for details). 
The pipeline reconstructs datacubes with 
a field of view of 1.125 $\times$1.150" from bi-dimentional raw frames. 
Telluric absorptions features were calibrated and removed based on the observations
of B5V stars. 
The wavelength calibration based on arc lamp spectra has an
accuracy of about 30\,\kms. 
A value of $V_{0}$=15.2\,km\,s$^{-1}$ was adopted as
rest velocity of \ots44 throughout the paper, 
which is the average radial velocity of T~Tauri stars and brown dwarfs 
in Cha\,I (Joergens 2006). 

{\bf Optical spectroscopy.}
An optical spectrum covering the range 0.5-1\,$\mu$m with a resolution
R=900 was taken by Luhman (2007)
with IMACS at the Magellan\,I telescope 
on January~6, 2005. 
This spectrum is used to analyze the H$\alpha$ emission of \ots44
and to provide an additional constraint to the photosphere model fit.
We calculate the heliocentric radial velocities relative to $V_{0}$.

%__________________________________________________________________
\section{Photospheric properties of OTS\,44} 

We find that it is not possible to fit the SED of \ots44 
when using the properties reported for this brown dwarf in the literature
($T_{\rm{eff}}$=2300\,K, $L_{*}$=0.00077\,L$_{\odot}$, $A_V$=0, Luhman 2007). 
Therefore, we performed a thorough modeling of the photosphere of \ots44 by applying 
the BT-Settl models (Allard et al. 2012). 
These models incorporate a sophisticated treatment of photospheric dust, 
which is likely to affect the cool atmosphere of \ots44.
We use broad-band photometry, a narrow grid of flux points calculated from 
the optical spectrum (Luhman 2007), 
and a surface gravity $\log(g)$ of 3.5, as determined from gravity-sensitive lines (Bonnefoy et al. 2013). 
We derive an effective temperature of $T_{\rm{eff}}$=1700\,K, a luminosity 
of $L_{*}$=0.0024\,$L_{\odot}$, and an extinction of $A_V$=2.6\,mag (Table\,\ref{tab:param}). 
Bonnefoy et al. (2013) previously found indications for a lower $T_{\rm{eff}}$ of \ots44. 
The probability distributions of the parameter values in our modeling approach 
are relatively broad, which hints at a remaining descrepancy with the models.
We conclude that the photospheric properties of \ots44 may still need to be refined in the future.

% -----------------------------------------------------------------
\begin{table}
\begin{center}
\caption{
\label{tab:param} 
Photospheric properties of \ots44.
}
\renewcommand{\footnoterule}{}  
\begin{tabular}{cccccc}		
\hline\hline  
\myrule

     & $T_{\rm{eff}}$ & $L_*$      & $A_V$ & $\log(g)$ & Reference  \\[0.15cm]
     &    [K]       &  [$\lsun$] & [mag] &        &    \\[0.15cm]
\hline 
\myrule
{\small Set\,1} & 2300 & 0.00077 & 0.0 &     & {\small Luhman 2007}  \\
{\small Set\,2} & 1700 & 0.0024  & 2.6 & 3.5 & {\small this work}    \\
\hline
\end{tabular}
\tablefoot{The errors for set\,2 are 
$\Delta$$T_{\rm{eff}}$=140\,K, $\Delta$$L_*$=0.00054\,$\lsun$, and $A_V<$3.0
with 90\% probability.
$\log(g)$ is taken from Bonnefoy et al. (2013).
$T_{\rm{eff}}$ in set\,2 agrees with the value found by
Bonnefoy et al. (2013), while $L_*$ slightly deviates between these two works. 
}
\end{center}
\end{table}

%__________________________________________________________________
\section{SED modeling of the disk of OTS\,44} 
\label{sect:sed}

We model the SED of \ots44 using the radiative transfer code \texttt{MC3D} (Wolf 2003) 
to characterize its circumstellar environment.
Because the disk parameters are strongly degenerate in the 
fitting procedure, we employ a passive-disk model 
consisting of a central substellar source 
surrounded by a parametrized disk. 

\emph{Dust distribution in the disk.}
We introduce a parametrized flared disk in which dust and gas are well mixed and homogeneous throughout the system.
This model has been successfully used to explain the observed SEDs of a large sample of young stars and 
brown dwarfs (e.g., Wolf et~al. 2003; 
Harvey et al. 2012).
For the dust in the disk we assume a density structure with a Gaussian vertical profile
$\rho_{\rm{dust}}=\rho_{0} (R_{*}/\varpi)^{\alpha} \exp( -z^2 / 2 h^2(\varpi)) $,
and a power-law distribution for the surface density 
$\Sigma(\varpi)=\Sigma_{0} (R_{*}/ \varpi)^p$,
where $\varpi$ is the radial distance from the central star measured 
in the disk midplane, and $h(\varpi)$ is the scale 
height of the disk.  
The outer disk radius $R_{out}$ is set to 100\,AU.
To allow flaring, the scale height follows the power law
$h(\varpi)$=$h_{100} (\varpi/100\,\rm{AU})^\beta$,
with the flaring exponent $\beta$ describing the extent of flaring and the 
scale height $h_{100}$ 
at $R_{out}$.

\emph{Dust properties.}
We consider the dust grains to be homogeneous spheres, which is a valid 
approximation to describe the scattering behavior as compared to a more complex description with fractal grain 
structures. The dust grain ensemble incorporates both astronomical silicate (62.5\%) and 
graphite (37.5\%) material. 
The grain size distribution is given by the standard power law $n(a)\propto{a^{-3.5}}$ 
with minimum and maximum grain sizes of 0.005\,$\mu{\rm m}$ and 0.25\,$\mu{\rm m}$, respectively.

\emph{Heating sources.}
We consider a passive disk 
with only stellar irradiation, but no viscous heating (e.g., Chiang \& Goldreich 1997).
Radiation heating of the dust from the accretion luminosity can be neglected
because $L_{acc}<0.2\%\,L_*$ (Sect.\,\ref{sect:emission}).
For $T_{\rm{eff}}$ and $L_{*}$ of the central source, we 
use the values derived here
(Table\,\ref{tab:param}, set~2) and d=162.5\,pc.
As said, an SED model using the parameters from
Luhman (2007; set~1 in Table\,\ref{tab:param}) cannot reproduce the observations.
The incident substellar spectrum is taken from the BT-Settl atmosphere database 
with $\log(g)$=3.5 
(Allard et al. 2012). 
The radiative transfer problem is solved self-consistently 
considering 100 wavelengths, logarithmically distributed in the range of 
[$0.05\,\mu{\rm{m}}$, $2000\,\mu{\rm{m}}$].

\begin{figure}[t]
  \centering
   \includegraphics[width=0.35\textwidth]{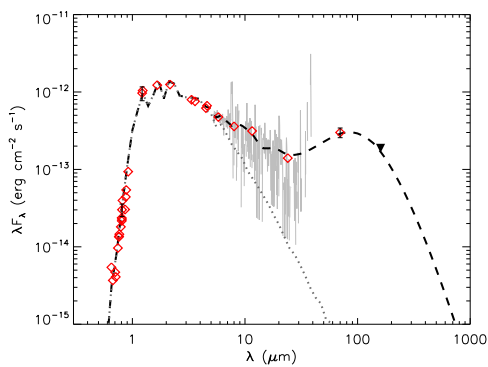}   
   \caption{SED of \ots44. Shown are photometric measurements (red diamonds) 
with errors if larger than the symbol, 
an upper limit for the 160\,$\mu{\rm{m}}$ flux (black triangle),
the mid-IR Spitzer/IRS spectrum (light gray), 
            the best-fit SED model (dashed line), and 
the input BT-Settl photosphere model 
(gray dotted line).}
   \label{best-fit}
\label{fig:bestfit}
\end{figure}

\emph{Fitting results.}
The SED fitting is performed with a hybrid strategy that combines the database method 
and the simulated annealing (SA) algorithm (Liu et al. 2013).
We first run a large grid of disk models with a broad range of 
disk parameters. Then SA is used to improve upon the results returned by the model grid
and to calculate local confidence intervals. 
The best-fit model is shown in Fig.\,\ref{fig:bestfit} and 
the corresponding disk parameter values are listed in Table\,\ref{tab:bestfit}. 
The best-fit model is not a unique solution due to model degeneracies between different 
parameters, for example $m_{\rm disk}$/$R_{out}$. We therefore
conduct a Bayesian analysis to estimate the validity range 
for each parameter (Pinte et al. 2008). 
We find that the best-fit and the most probable values 
(cf. Sect.\,\ref{sect:append})
agree well with each other, 
in particular the disk mass,
which demonstrates that the modeling efforts place good constraints on the mass and structure 
of the disk of \ots44. 
We note that strong grain growth would remain undetected in our data ($\leq$160\,$\mu m$) and
could affect $m_{\rm disk}$.
The values of most disk parameters derived here are largely consistent
with the results of Harvey et al. (2012, 
$R_{\rm in}$=0.01\,AU, $m_{\rm disk}$= $5.0\times10^{-5}\,\msun$,
$\beta$=1.15,
$h_{\rm 100}$=15\,AU, $i$= 60$^{\circ}$),
who used a coarser grid resolution and slighlty different dust properties 
and photospheric parameters. 
An SED model by Bonnefoy et al. (2013) aimed at understanding whether the disk could create a near-IR
excess that would bias the temperature estimate showed that this is not the case.

%__________________________________________________________________
\section{Paschen\,$\beta$ and H$\alpha$ emission} 
\label{sect:emission}

\begin{figure}[t]
\centering
\includegraphics[width=.55\linewidth,clip]{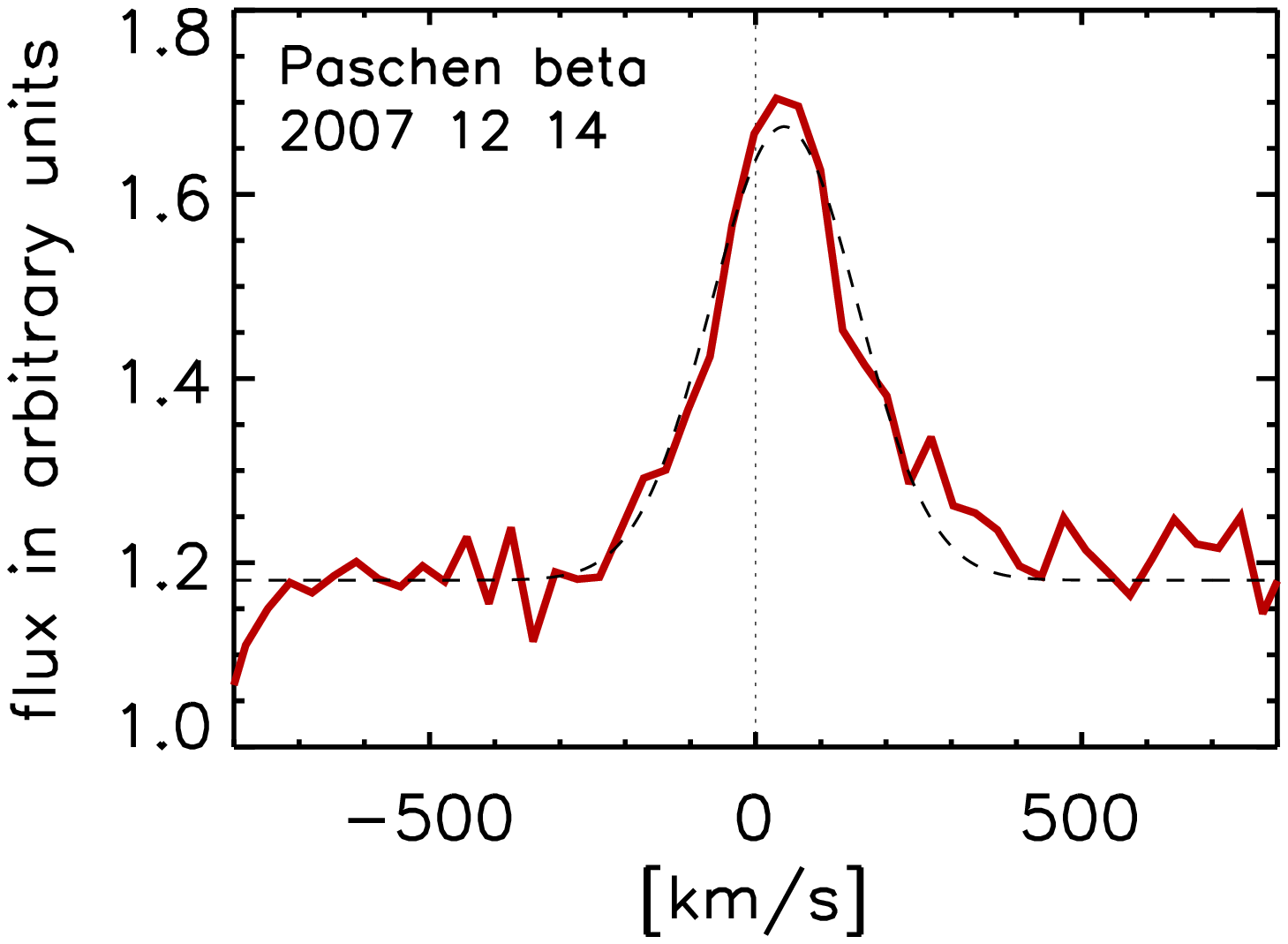}
\includegraphics[width=.55\linewidth,clip]{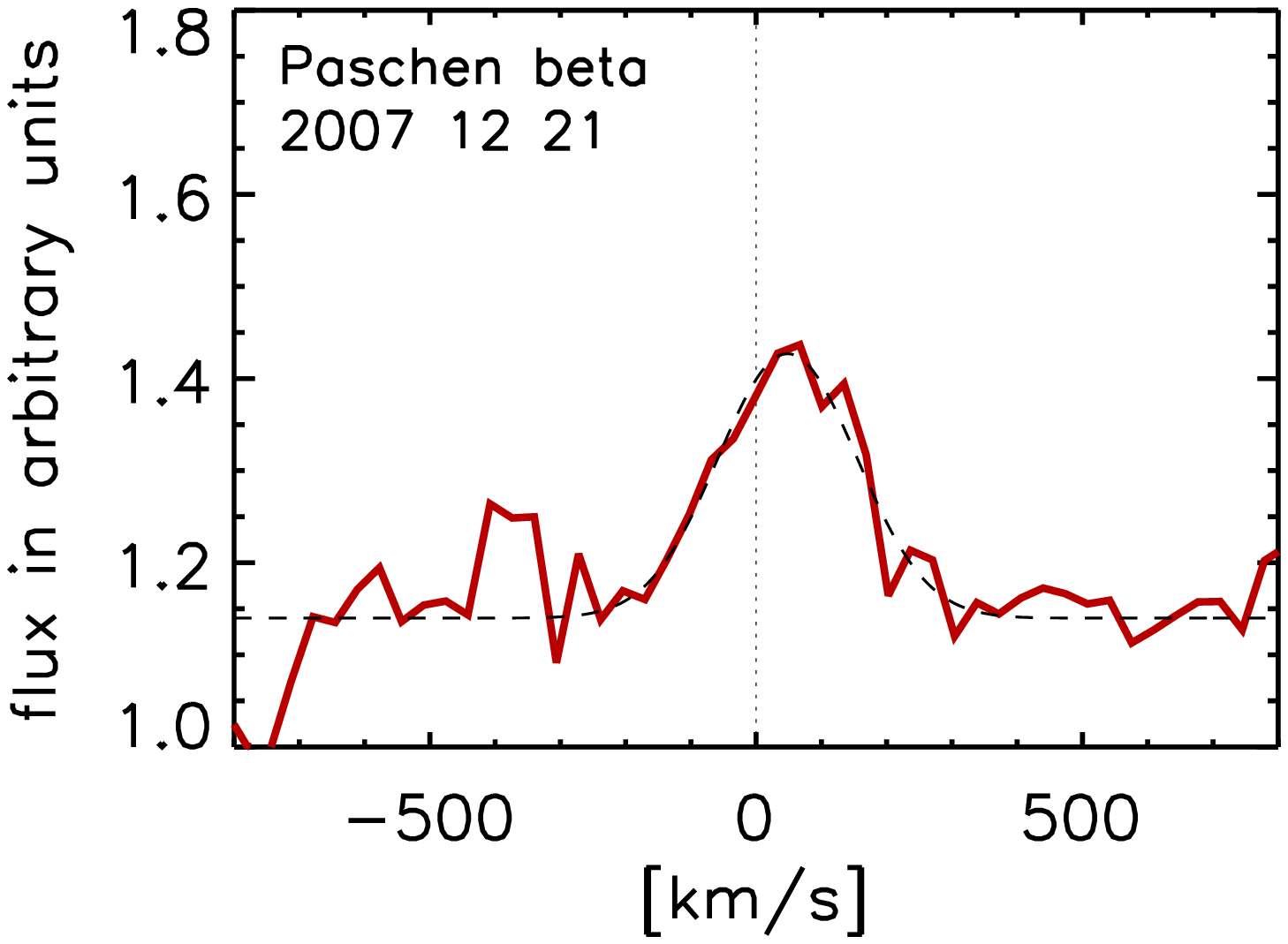}
\includegraphics[width=.55\linewidth,clip]{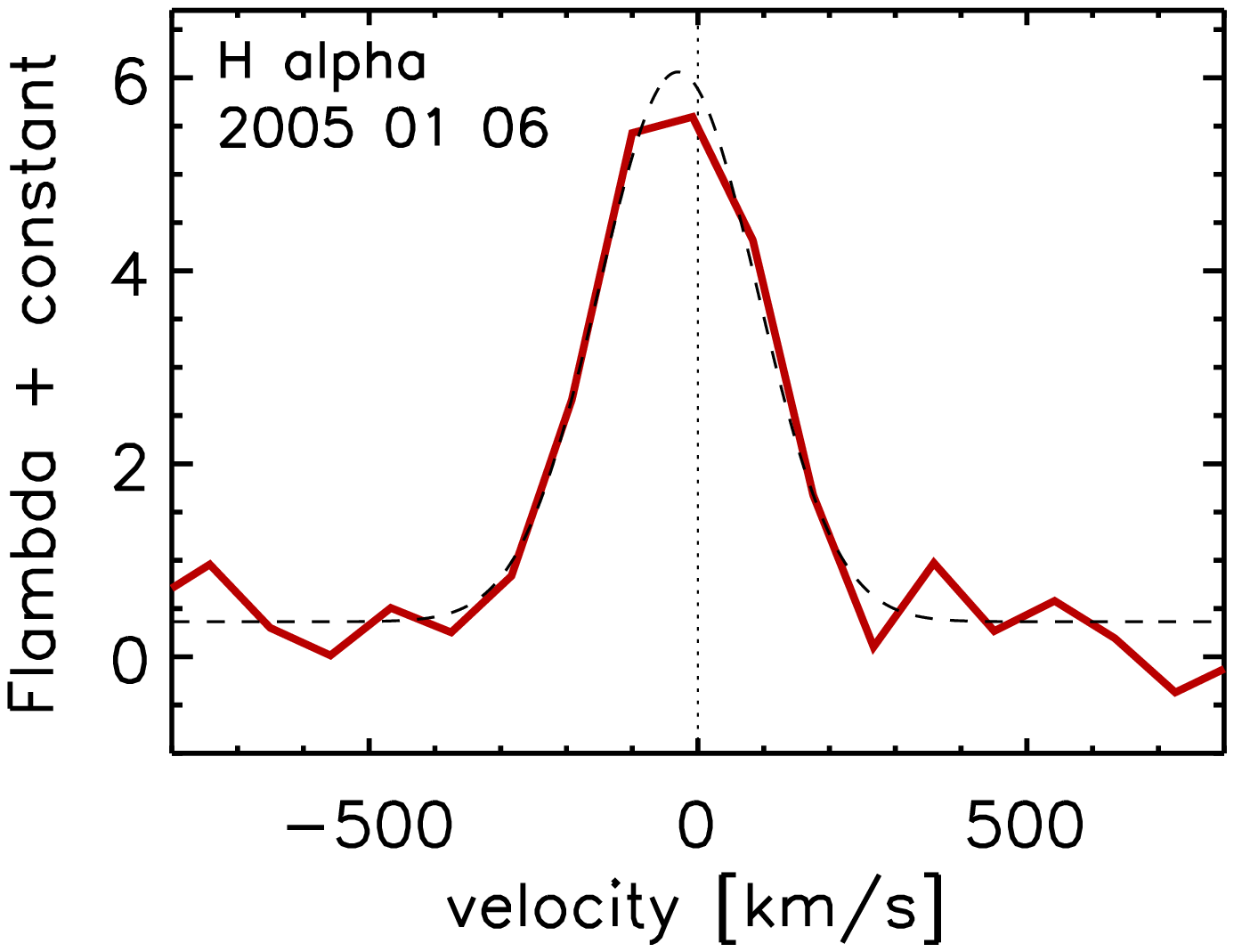}
\caption{
\label{fig:pabeta}
Pa\,$\beta$ emission of \ots44 in SINFONI/VLT spectra 
and H$\alpha$ emission of \ots44 based on a spectrum from Luhman (2007).
The dashed lines are Gaussian fits to the profiles.
}
\end{figure}

We discover a strong and broad Paschen\,$\beta$ (Pa\,$\beta$) emission line in our near-IR SINFONI spectra
of \ots44 (Fig.\,\ref{fig:pabeta}). Furthermore, a prominent H$\alpha$ emission line is visible
in the optical spectrum of Luhman (2007, see Fig.\,\ref{fig:pabeta}). 
Both of these Hydrogen emission lines exhibit a broad profile 
with velocities of $\pm$200\,\kms or more.
We investigate the properties and origin of these lines through a line profile analysis. 
We determine the equivalent width (EW)
by directly integrating the flux within the line region and the EW errors
following Sembach \& Savage (1992). Furthermore, we 
measure the line center and full-width-at-half-maximum (FWHM) based on a 
Gaussian fit to the profiles.
Table\,\ref{tab:ew} lists the results.

The H$\alpha$ line has a symmetrically shaped profile with an EW of -141\,{\AA}, 
demonstrating that \ots44 is actively accreting
(e.g. Barrado \& Martin 2003).
The line appears to be blueshifted with its center located at -30\,\kms.
Higher resolution spectroscopy is needed to determine whether
this shift is real.

The shape of the Pa\,$\beta$ line appears to be slightly asymmetric, with the red wing being more pronounced. 
The profile is significantly variable between the two observing epochs separated by a few days.
We measure an EW of -6.7 and -4.2\,{\AA} for the spectra from December 14 and 
21, respectively. 
The line has a peak at redshifted velocities at about 40-50\,\kms.
We see a redshift of similar order also in photospheric lines of \ots44.
While the Pa\,$\beta$ emission of T~Tauri stars is mostly attributed to magnetospheric 
accretion and winds (e.g., 
Rigliaco et al. 2012), there is observational evidence that 
part of the Pa\,$\beta$ emission, in particular the broad line wings,
can be formed by other processes, such as outflows (e.g., 
Whelan et al. 2004).
We conducted a spectro-astrometric analysis of the Pa\,$\beta$ line 
in the SINFONI 3D cube data to locate the formation site of this emission.
We find no spectro-astrometric signal in this line that exceeds 5\,mas (0.8\,AU). 

{\em Mass accretion rate.}
We determine the mass accretion rate of \ots44 based on the H$\alpha$ line by 
assuming that the H$\alpha$ emission is entirely formed by accretion processes.
For this purpose, we calculate the H$\alpha$ line luminosity from the H$\alpha$ EW using broad-band photometry
and converte it into an accretion luminosity by applying the empirical relation of 
Fang et al. (2009, cf. also Joergens et al. 2012b).
We use a dereddened $R$-band magnitude, a distance of 162.5\,pc, 
the set~2 parameters (Table\,\ref{tab:param}) for the extinction and for calculating a radius 
via the Stefan-Boltzmann law, 
and a mass of 11.5\,$\mjup$, which is an intermediate value of the estimated range 
of 6-17\,$\mjup$ (Luhman et al. 2005b; Bonnefoy et al. 2013).
We derive an H$\alpha$ line luminosity of 
6.9$^{+1.4}_{-1.2}\times$10$^{-7}\lsun$,
an accretion luminosity of 
3.7$^{+17.4}_{-3.1}\times$10$^{-6}\lsun$
(0.15\% of L$_*$),
and a mass accretion rate of
7.6$^{+36}_{-6.4}\times 10^{-12}$\,$\mperyr$. We take into account 
errors in EW, R$_{\rm mag}$, and in the empirical relation.
We furthermore attempt to estimate the mass accretion rate based on the 
Pa\,$\beta$ line by using the empirical relation of Rigliaco et al. (2012),
a dereddened $J$-band magnitude, and again set~2 of Table\,\ref{tab:param}.
This yields a mass rate that is two orders of magnitude higher than that derived 
from H$\alpha$, namely 1.7$\times 10^{-9}$\,$\mperyr$
for epoch 1 (December 14) and 8.5$\times 10^{-10}$\,$\mperyr$ for epoch 2 (December 21).
This descrepancy
lets us speculate 
that part of the Pa\,$\beta$ line might come from a different origin than accretion.

% ----------------------------------------------------------------------
\begin{table}[t]
 \centering
   \caption{Disk parameter values of the best-fit SED model.}
    \begin{tabular}{cccccccc}
     \hline
     \hline
\myrule
$R_{\rm in}$ & $R_{\rm out}$  & $p$ & $\beta$ & $h_{\rm 100}$ & $m_{\rm disk}$ & $m_{\rm disk}$ & $i$ \\ 
AU] & [AU] & &  & [AU] & [10$^{-5}\,\rm M_{\odot}$] & [$\rm M_{\oplus}$] & [$^{\circ}$] \\ 
\hline
\myrule
0.023 & 100 & 1.136 & 1.317 & 17.42 & 9.06 & 30.2 & 58 \\
   \hline 
   \end{tabular}
\tablefoot{
For the photosphere, the values of set~2 of Table\,\ref{tab:param} were used.
See Sect.\,\ref{sect:append} for confidence intervals of the disk parameters.
}
\label{tab:bestfit}
\end{table}

% -----------------------------------------------------------------
\begin{table}
\begin{center}
\caption{
\label{tab:ew} 
Observed emission lines of \ots44.
}
\renewcommand{\footnoterule}{}  
\begin{tabular}{ccccc}		
\hline\hline  
\myrule
line & date & $V_{center}$ & FWHM   & EW                \\[0.15cm]
     &      & [km/s]      & [km/s] & [\r{A}]           \\[0.15cm]
\hline 
\myrule
H\,$\alpha$       & 2005 01 06 & -30       & 283       & -141 $\pm$ 14 \\
Pa\,$\beta$       & 2007 12 14 & +44 & 265 & -6.7 $\pm$ 0.3 \\
Pa\,$\beta$       & 2007 12 21 & +48 & 251 & -4.2 $\pm$ 0.3 \\
\hline
\end{tabular}
\end{center}
\end{table}

% ------------------ Pa beta + Halpha  -----------------------------------------------
% -----------------------------------------------------------------

% ------------------ Mdisk-mass plot  -----------------------------------------------
\begin{figure}[t]
\centering
\includegraphics[width=\linewidth]{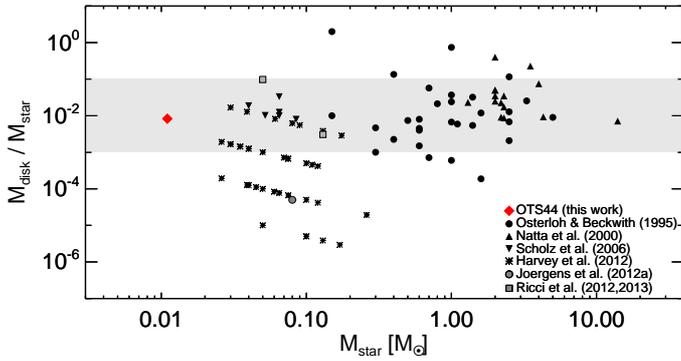}
\caption{
\label{fig:mdisk}
Relative disk mass versus central mass of stars and brown dwarfs including 
\ots44 (red diamond). 
We note that 
for $\rho$\,Oph102 we use $M_*$=0.13\,$\msun$ 
(M5.5, K.\,Luhman, pers.\,comm.), different from Ricci et al. (2012, 0.06\,$\msun$).
}
\end{figure}
% -----------------------------------------------------------------

% ------------------ Macc-mass plot  -----------------------------------------------
\begin{figure}[t]
\centering
\includegraphics[width=\linewidth]{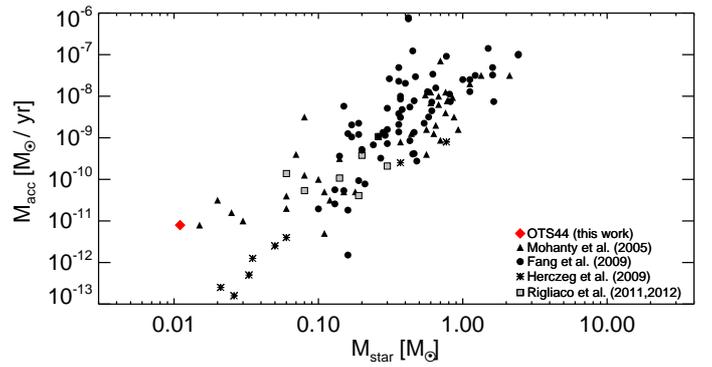}
\caption{
\label{fig:macc}
Mass accretion rate versus central mass of stars and brown dwarfs
including \ots44 (red diamond). 
}
\end{figure}
% -----------------------------------------------------------------

\section{Conclusions}

We have discovered strong, broad, and variable Pa\,$\beta$ emission of 
the young very low-mass brown dwarf \ots44 (M9.5)
in VLT/SINFONI spectra, which is evidence for active accretion at the planetary border.  
We determined the properties of the disk that surrounds \ots44 through 
\texttt{MC3D} radiative transfer modeling of flux measurements from the optical to the far-IR (Herschel).
We found that \ots44 has a highly flared disk ($\beta>$1.2) with a mass of 
9.1\,$^{+1.7}_{-5.5}\times10^{-5}\,\msun $,
that is about 0.1\,$\mjup$ or $30\,M_{Earth}$. 
We also investigated the H$\alpha$ line of \ots44 in a spectrum from Luhman (2007) and 
found strong H$\alpha$ emission with an EW of -141\,{\AA} indicative of active accretion.
Both the Pa\,$\beta$ and H$\alpha$ emission lines of \ots44 have broad profiles 
with the wings extending to velocities of about $\pm$200\,\kms.
The Pa\,$\beta$ emission is significantly variable on timescales of a few days, 
indicating variability in accretion-related processes of \ots44.
We estimated the mass accretion rate of \ots44 to 
7.6$^{+36}_{-6.4}\times 10^{-12}$\,$\mperyr$
by using the H$\alpha$ line.
A mass accretion rate based on the Pa\,$\beta$ line gives a significantly higher value, and 
we speculate that part of the Pa\,$\beta$ emission might come from other processes related 
to accretion, such as outflows.
Furthermore, in the course of studying \ots44, we fitted a photospheric BT-Settl model to its
optical and near-IR SED and derived a lower effective temperature and higher extinction than was 
previously found (Luhman 2007).

We have presented the first detection of Pa\,$\beta$ emission for an object at the deuterium-burning limit.
Our analysis of Pa\,$\beta$ and H$\alpha$ emission of \ots44 demonstrates that objects of a few Jupiter masses 
can be active accretors.
Furthermore, \ots44 is the lowest-mass object to date for which the disk mass is determined based on far-IR data.
Our detections therefore extend the exploration of disks and accretion during the T~Tauri phase   
down to the planetary mass regime.
Plotting the relative disk masses of stars and brown dwarfs 
including \ots44 (Fig.\,\ref{fig:mdisk}) shows
that the ratio of the disk-to-central-mass of about 10$^{-2}$ found for 
objects between 0.03\,$\msun$ and 14\,$\msun$ is also valid for \ots44 at a mass of about 0.01\,$\msun$.
Furthermore,
the mass accretion rate of \ots44 is consistent with a decreasing
trend from stars of several solar masses to brown dwarfs down to 0.01\,$\msun$ (Fig.\,\ref{fig:macc}). 
It is also obvious from this figure that \ots44
has a relatively high mass accretion rate considering its small mass.
These observations show that the processes that accompany canonical star formation,
disks and accretion, are present down to a central mass of a few Jupiter masses. 
\ots44 plays a key role in the study of disk evolution and accretion 
at an extremely low mass and, therefore, in constraining the minimum mass that star formation can produce.
It will be the target of our future observational efforts.

\begin{acknowledgements}
We thank K. Luhman for providing the optical spectrum of \ots44 and related information,
C. Dumas and A.-M. Lagrange for help with the SINFONI data reduction,
the ESO staff at Paranal for executing the SINFONI observations
in service mode, and the anonymous referee for very helpful comments. 
This work made use of the NIST database (Kramida et al. 2012) and VOSA (Bayo et al. 2008).
MB acknowledges funding from Agence Nationale pour la Recherche, 
France through grant ANR10-BLANC0504-01.
\end{acknowledgements}

\Online

% -------------------------------------------- appendix ------------------------
\begin{appendix} 
\section{Bayesian probability analysis}
\label{sect:append}

\begin{figure*}[h]
\centering
\hbox{\mbox{}
   \hfill\includegraphics[width=0.2\textwidth,clip]{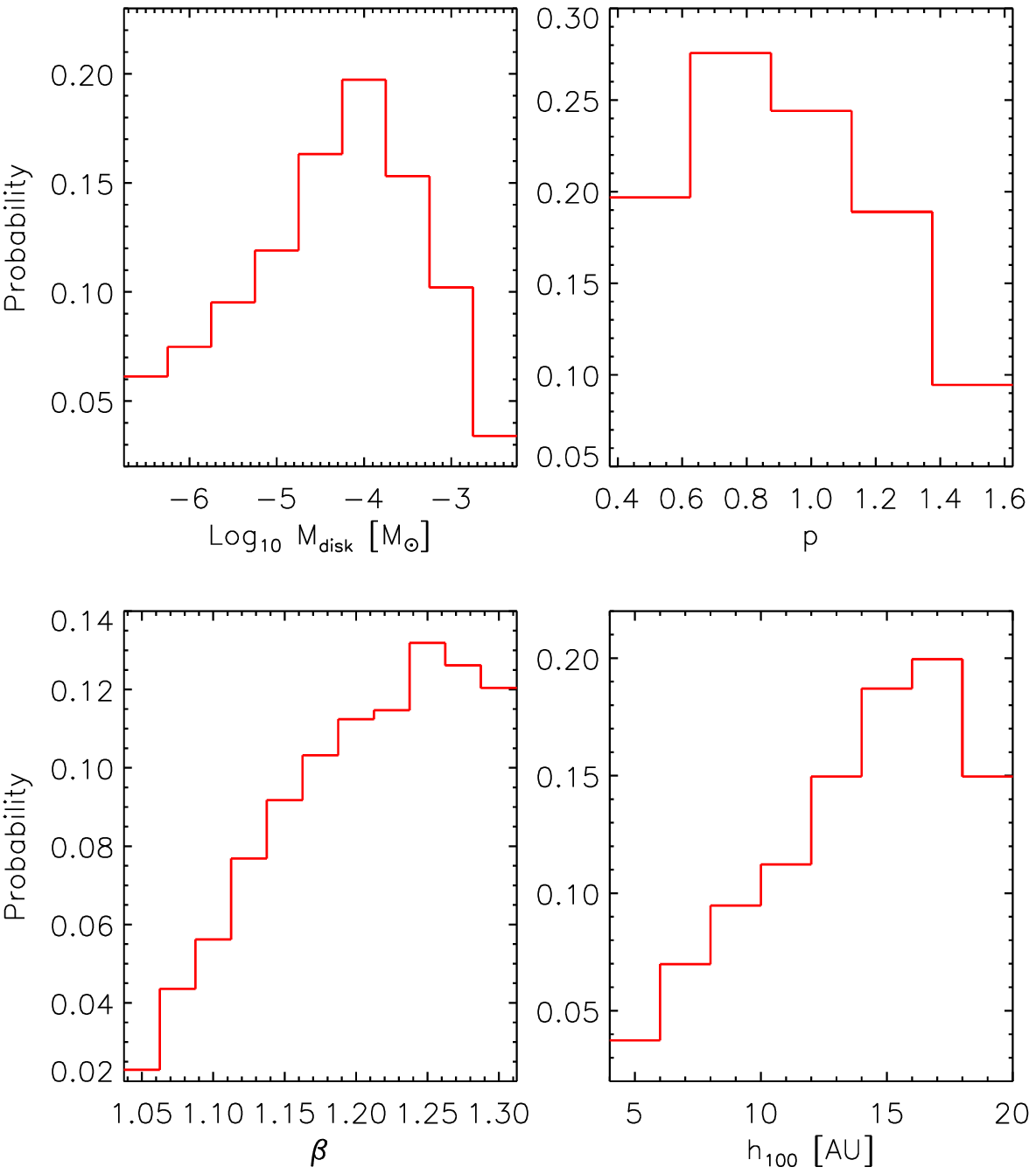}   
   \hfill\includegraphics[width=0.187\textwidth,clip]{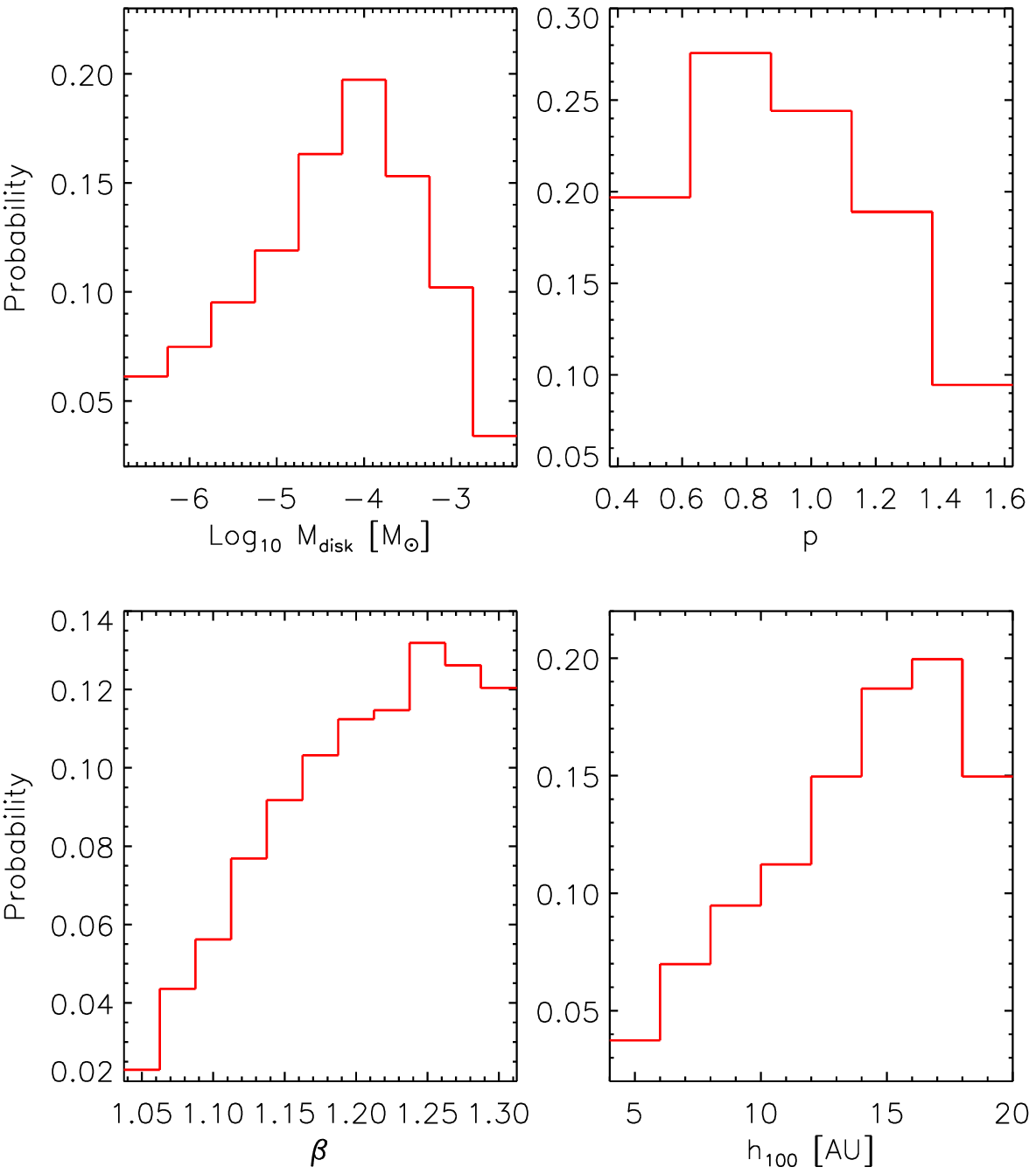} 
   \hfill\includegraphics[width=0.182\textwidth,clip]{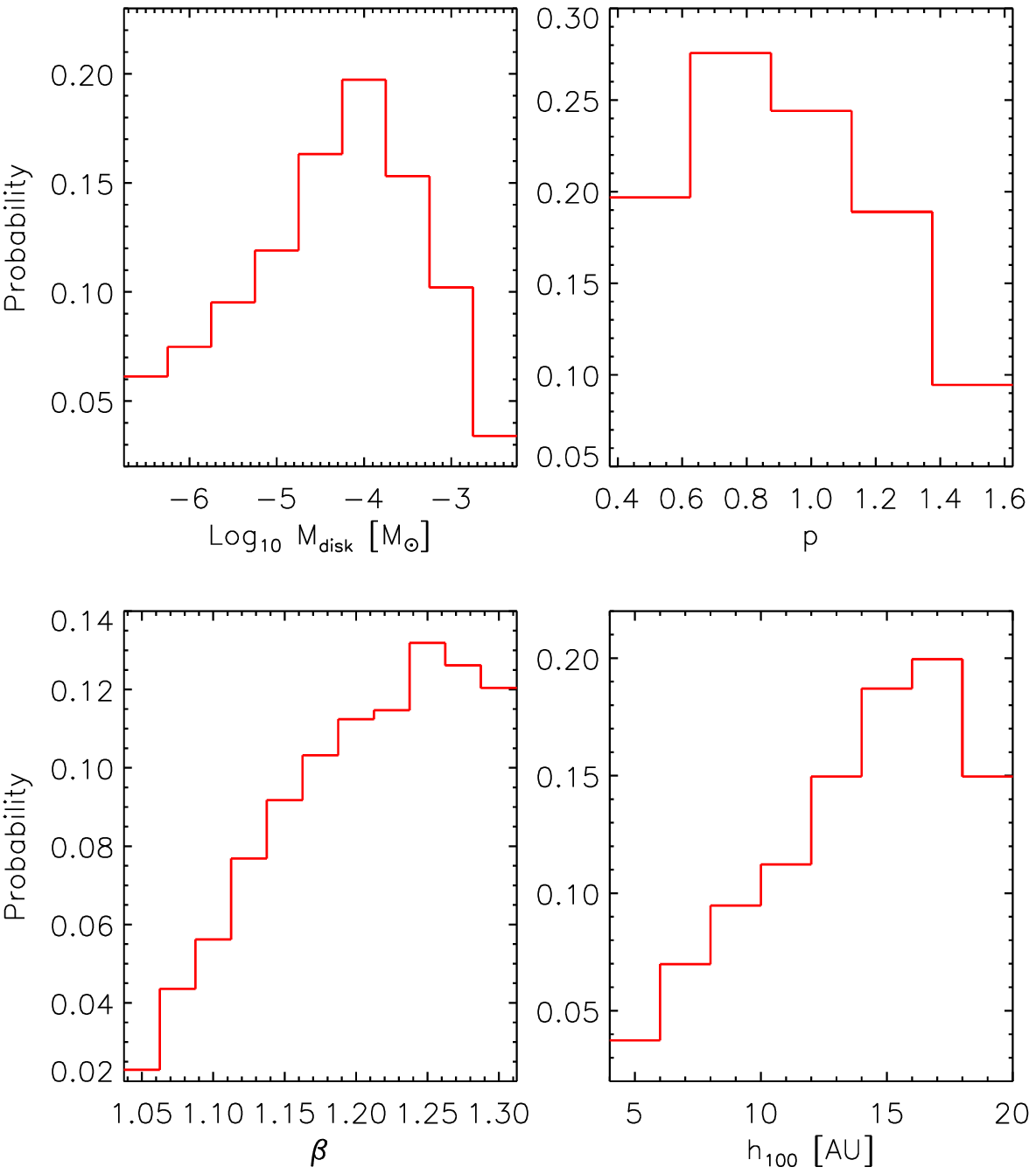}   
   \hfill\includegraphics[width=0.185\textwidth,clip]{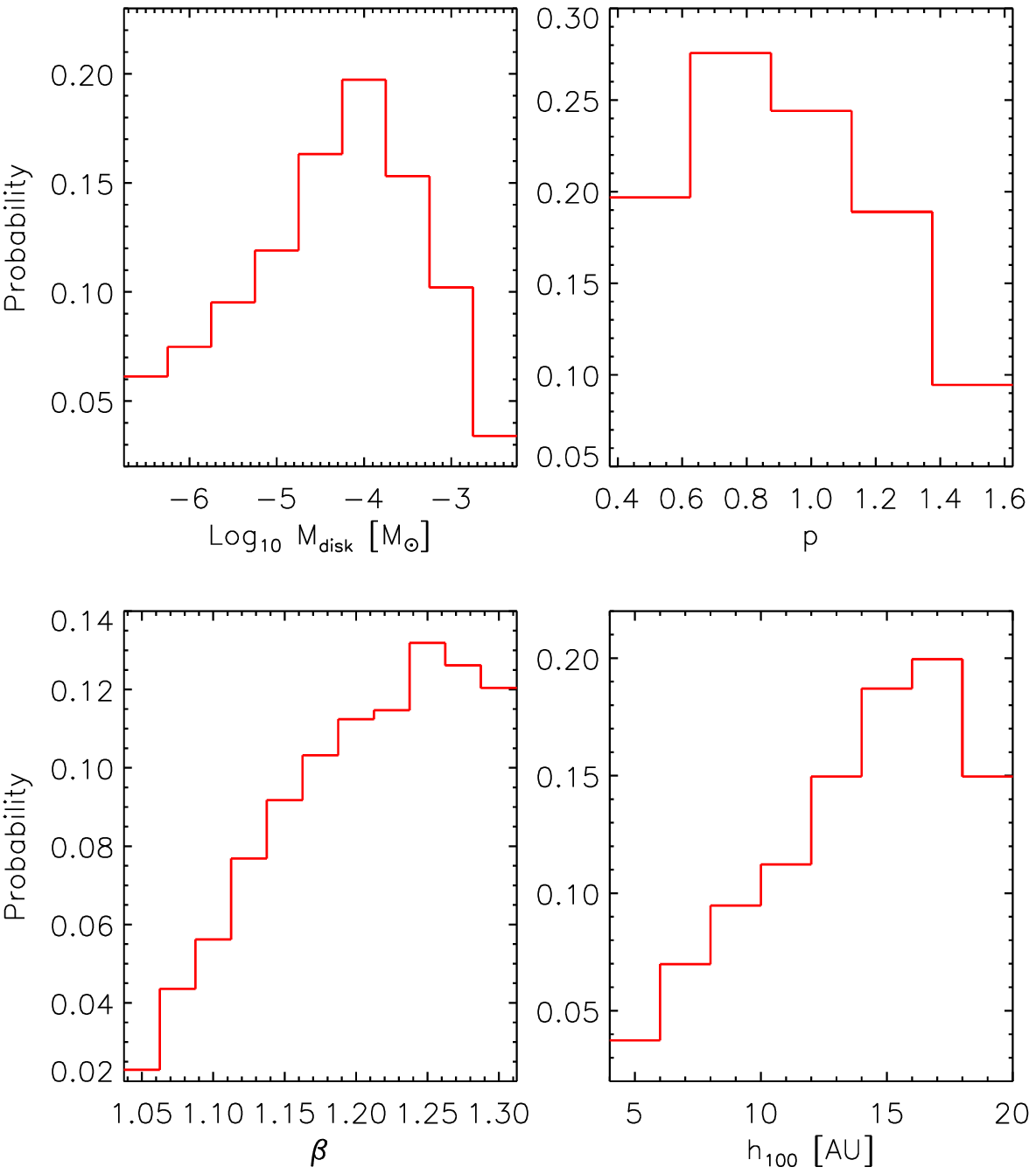} 
   \hfill\includegraphics[width=0.18\textwidth,height=0.195\textwidth,clip]{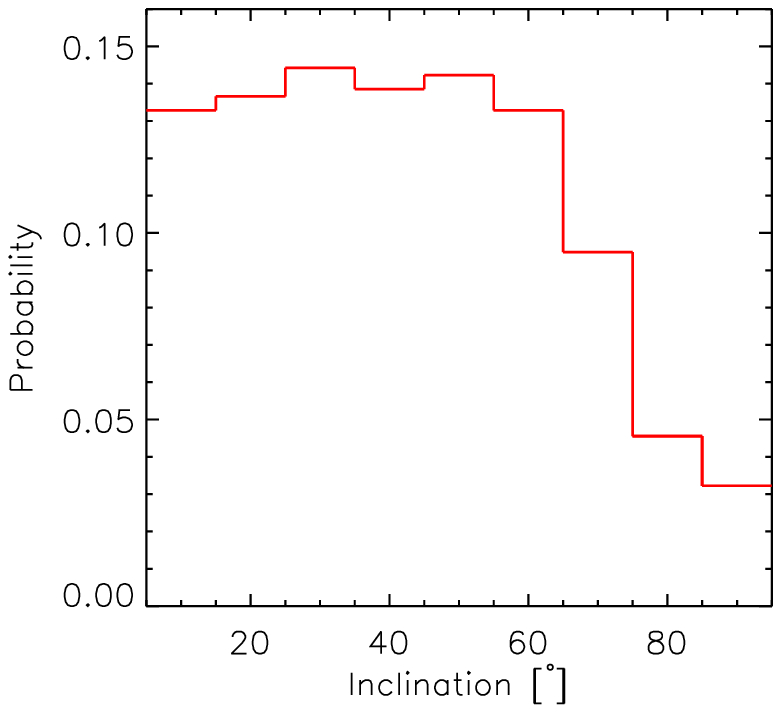}  
}
   \caption{Bayesian probability distributions of selected disk parameters.}
%   \label{bayesian_prob}
\label{fig:bayesian_prob}
\end{figure*}

% ----------------------------------------------------------------------
\begin{table}[h]
 \centering
   \caption{Confidence intervals of the disk parameter values of \ots44.}
    \begin{tabular}{lcc}
     \hline
     \hline
\myrule
Parameter  &  best model & valid range    \\
\hline
\myrule
$R_{\rm in}$ [AU]  & 0.023 $^{+0.018}_{-0.013}$ & 0.01-0.04 \\ 
\myrule
$R_{\rm out}$ [AU] & 100                      & fixed     \\
\myrule
$p$               & 1.136 $^{+0.051}_{-0.025}$ & 0.62-1.27 \\
\myrule
$\beta$           & 1.317 $^{+0.017}_{-0.042}$ & 1.16-1.32 \\
\myrule
$h_{\rm 100}$ [AU] & 17.42 $^{+0.68}_{-2.55}$   & 9.0-18.5  \\
\myrule
$m_{\rm disk}$ [10$^{-5}\,\rm M_{\odot}$] & 9.06 $^{+1.72}_{-5.46}$ & 0.32-56.2 \\
\myrule
$i$ [$^{\circ}$]   & 58$^{+6}_{-9}$ & 18-65 \\
%\myrule
   \hline 
   \end{tabular}
\tablefoot{
Parameter values of the best-fit SED model with local errors
and the valid ranges for each parameter. 
The local error is deduced based on simulated annealing 
by probing the direct environment of the best fit with a Markov chain.
The valid range for each parameter gives a 68\% confidence interval based on 
the Bayesian probability distribution, as shown in Fig.\,\ref{fig:bayesian_prob}. 
}
\label{tab:intervals}
\end{table}

\end{appendix}
\end{document}